\documentclass[printer]{aa}
\usepackage[varg]{txfonts}
\usepackage{graphicx}
\usepackage{fixltx2e}
\usepackage{wasysym}

\bibpunct{(}{)}{;}{a}{}{,} 

\begin{document}

\title{Damping of nonlinear standing kink oscillations: a numerical study}

\author{N. Magyar
\and T. Van Doorsselaere }

\institute{Centre for mathematical Plasma Astrophysics, Department of Mathematics, KU Leuven, \mbox{Celestijnenlaan} 200B, bus 2400,
3001 Leuven, Belgium}

\abstract{}{We aim to study the standing fundamental kink mode of coronal loops in the nonlinear regime, investigating the changes in energy evolution in the cross-section and oscillation amplitude  of the loop which are related to nonlinear effects, in particular to the development of the Kelvin-Helmholtz instability (KHI). } {We run ideal, high-resolution three-dimensional (3D) magnetohdydrodynamics (MHD) simulations, studying the influence of the initial velocity amplitude and the inhomogeneous layer thickness. We model the coronal loop as a straight, homogeneous magnetic flux tube with an outer inhomogeneous layer, embedded in a straight, homogeneous magnetic field. } {We find that, for low amplitudes which do not allow for the KHI to develop during the simulated time, the damping time agrees with the theory of resonant absorption. However, for higher amplitudes, the presence of KHI around the oscillating loop can alter the loop's evolution, resulting in a significantly faster damping than predicted by the linear theory in some cases. This questions the accuracy of seismological methods applied to observed damping profiles, based on linear theory.} {}

\titlerunning{Damping of nonlinear kink oscillations}
\maketitle

\section{Introduction}

With the first observations of transverse oscillations in coronal loops by TRACE \citep{1999ApJ...520..880A,1999Sci...285..862N} a new era has begun in the exploration and understanding of the solar corona. These observations of waves allow the inference of different physical parameters which were previously largely unknown, a method called coronal seismology. It was quickly noted that the observed coronal loop oscillations were strongly damped, and if such a high dissipation was due to viscous or resistive damping, then the associated transport coefficients must be orders of magnitude higher than predicted by the classical theory. Nearly all of the observed, high-amplitude, lower coronal eruption-related coronal loop oscillations \citep{2015A&A...577A...4Z} are strongly damped, with rare but nevertheless very intriguing exceptions (see, e.g. \citet{2002SoPh..206...99A, 2011ApJ...736..102A,2012ApJ...751L..27W}). Recently, small amplitude decayless kink oscillations  were observed in coronal loops \citep{2013A&A...552A..57N,2013A&A...560A.107A}, and are thought to be ubiquitous in active regions \citep{2015A&A...583A.136A}. These oscillations appear to be constantly driven, and do not originate from an impulsive or eruptive event like the high amplitude oscillations. It is generally accepted that the mechanism responsible for the fast damping of transverse oscillations is the extensively studied resonant absorption or mode coupling \citep[e.g.][]{1978ApJ...226..650I,1984ApJ...277..392H,2002ApJ...577..475R,2002A&A...394L..39G}. For resonant absorption to damp the oscillations, the existence of a surface within the limits of the loop, with Alfvén speed matching the global kink phase speed, is required. This is usually portrayed as a smooth layer around a homogeneous flux tube, but resonance has been proven to exist regardless of the geometrical shape of such a layer \citep{2008ApJ...679.1611T,2011ApJ...731...73P}. The linear theory of kink oscillations is well studied \citep[for a review, see][]{2009SSRv..149..199R}. The damping profile of oscillations, in the presence of a driver (or initial perturbation in the case of standing modes) has been shown to deviate from a purely exponential decay, being described rather by an initially Gaussian damping profile followed by an exponential damping profile \citep{2012A&A...539A..37P,2013A&A...551A..39H,2013A&A...551A..40P,2016A&A...585L...6P,2016A&A...589A.136P}. 
Observed average periods, amplitudes and exponential damping times of kink oscillations in coronal loops were reported by,  \citet{2002SoPh..206...99A}, for example. In their analysis of 26 loop oscillation events, they find average oscillation periods of $321 \pm 140\ \mathrm{s}$, oscillation amplitudes of $2200 \pm 2800\ \mathrm{km}$ and damping times of $580 \pm 385\ \mathrm{s}$. The measured amplitudes correspond to relative amplitudes (to the loop length $L$) of approximately 1-5\%. \citet{2007A&A...463..333A} and \citet{2008A&A...484..851G} used the analytical formula for the damping time \citep{2002ApJ...577..475R} to construct a seismological method for inferring the loop parameters from the observed damping time and period. 
\\Waves in the solar atmosphere often have high enough amplitudes to be considered nonlinear. The study of nonlinear waves in the solar atmosphere has been carried out especially in the context of chromospheric and coronal heating. For a review of earlier theoretical work on the subject, see \citet{2006RSPTA.364..485R}. In particular, the analytical theory of nonlinear kink oscillations has been studied, both for propagating \citep{2010PhPl...17h2108R}, and standing waves \citep{2014SoPh..289.1999R}. In these studies it was shown that damping of propagating kink waves can be enhanced by the nonlinearity of an $m$-mode resonance, where energy from the $m=1$ kink mode is transferred to $m \geq 2$ fluting modes, which, by resonant absorption, can damp faster  due to shorter wavelengths. It was noted that the $m$-mode resonance can damp the kink wave even in the absence of a resonant layer. However, in an axially inhomogeneous flux tube,  due to density stratification, for example, the $m$-modes no longer have the same phase speed, thus the $m$-mode resonance and enhanced damping disappears. These calculations were valid for weakly nonlinear oscillations, and the authors anticipated that the amplitude would be affected in the fully nonlinear case. In general, the study of fully nonlinear problems is only possible numerically.  Numerical studies of nonlinear kink oscillations of coronal loops have been carried out by, \citet{2004ApJ...610..523T,2005SSRv..120...67O,2008ApJ...687L.115T,2009ApJ...694..502O,
2014ApJ...787L..22A,2015ApJ...809...72A,2015A&A...582A.117M,2016arXiv160404078M}, for example.\ See also \citet{2009SSRv..149..153O} for a review. Of particular and renewed interest in nonlinear evolution of  transverse waves, even for low amplitudes, is the susceptibility of the resonant layer, due to its high velocity shear, to the Kelvin-Helmoltz instability (KHI) \citep{1983A&A...117..220H,1984A&A...131..283B,1994ApJ...421..372K,1994GeoRL..21.2259O,1997SoPh..172...45P}. This can enhance the wave energy dissipation via local turbulence, and thus is also relevant in the coronal heating problem. 
Direct observational evidence is lacking of the KHI in coronal loops. However, it has been observed in coronal mass ejections and quiescent prominences \citep{2010ApJ...716.1288B,2010SoPh..267...75R,2011ApJ...729L...8F,2011ApJ...734L..11O}, and it was suggested that, in fact, the KHI may appear as strands of coronal loops as seen in EUV \citep{2014ApJ...787L..22A}. The previous statement is strengthened by the first observational evidence of resonant absorption in prominences \citep{2015ApJ...809...71O,2015ApJ...809...72A}. Recently, the idea of observing the damping profile of kink oscillations of coronal loops in order to infer various local parameters such as density ratios and inhomogeneous layer widths in the context of coronal seismology was put forward. \citep{2016A&A...589A.136P}. Therefore, it is essential to know the effects that nonlinearity might have on damping profiles. New observational evidence suggests that the damping of the transverse oscillations of coronal loops is dependent on the amplitude \citep{2016A&A...590L...5G}. \\
In this paper we investigate the nonlinear standing kink oscillation of coronal loops, for which the main effect of nonlinearity is the development of the KHI at the loop edges, altering the energy distribution in the loop cross section and ultimately leading to a change in the damping profile of the loop displacement.

\section{Numerical Model}
 
 \subsection{Initial conditions}

We model the coronal loop as a straight, density-enhanced magnetic flux tube. The loop consists of a homogeneous inner core, and an inhomogeneous annulus (transitional layer), in which the density varies sinusoidally from the core ($\rho_i = 2.5 \cdot 10^{-12}$ kg m$^{-3}$) to the background density value ($\rho_e = 0.5 \cdot 10^{-12}$ kg m$^{-3}$), where the  subscripts $i,e$ stand for internal and external, respectively. The numerical domain is permeated by a straight, homogeneous magnetic field, of $12.5$ G strength. The distance between the central axis of the flux tube and the midpoint of the inhomogeneous layer defines its radius, $R = 1.5$ Mm. The thickness of the inhomogeneous layer is denoted by $l$. We neglect the effects of curvature, gravity (i.e. no stratification), energy sources and sinks (heating, thermal conduction, radiative processes), and we lack a realistic lower solar atmosphere (i.e. photosphere, chromosphere). The plasma-$\beta$ is constant throughout the domain, $0.06$, corresponding to $T_e = 4.5\ \mathrm{MK}$ and $T_i = 0.9\ \mathrm{MK}$. The resulting temperature profile is most probably not realistic. This simplistic profile was chosen in order to start with an initial equilibrium, that is, constant total pressure throughout the domain. Tests with isothermal models ($T_e = T_i = 0.9\ \mathrm{MK}$) show no significant difference in comparison with the present model, as expected (period, growth rates and damping rates not depending on the temperature profile). \\
Initially, we impose a perturbation in a component of the velocity transverse to the loop axis, of the form
\begin{equation}
V_y = \left\{
 \begin{array}{cl}
     A\ V_{\mathrm{A,i}}\ \mathrm{cos}\left(\frac{\pi z}{L}\right), &  x^2 + y^2 \leq    \left(R+\frac{l}{2}\right)^2\\
     0, & \mathrm{otherwise}
  \end{array}
  \right.
\end{equation}
 $V_\mathrm{{A,i}}$ is the Alfvén speed inside the loop ($V_{\mathrm{A,i}} \approx 0.7\ \mathrm{Mm\ s^{-1}}$), $L = 120$ Mm is the total loop length, and $A$ is the perturbation amplitude. Thus the perturbation acts only inside the loop, including the inhomogeneous layer. This excites the fundamental kink mode of the loop, and we stop the simulation at $t_\mathrm{f} = 1500$ s.

 \subsection{Boundary conditions}
 
In order to excite a standing transverse oscillation, we fix the footpoint of the loop ($z = 60$ Mm), by setting the transverse velocities to antisymmetric, while the other variables are set to continuous (Neumann-type, zero-gradient boundary condition) at this boundary. Exploiting the symmetric properties of the fundamental kink mode, we model only half of the loop in both the $z$ and $x$ directions. This reduces the computational time four-fold. The boundary conditions for these mirroring boundaries are described in \citet{2015A&A...582A.117M}. At the other, lateral boundaries we let any waves leave the domain freely by imposing an outflow (zero-gradient) condition on all variables. 

 \subsection{Numerical method and mesh}
 
To solve the ideal 3D MHD problem, we use the \texttt{MPI-AMRVAC} code \citep{2012JCoPh.231..718K,2014ApJS..214....4P}. We use the implemented second-order `onestep' TVD method with the Roe solver and `Woodward' slope limiter. The constraint on the magnetic field divergence is maintained using Powell's scheme. The numerical domain has dimensions of $(0,6) \times (-10,10) \times (0,60)$ Mm. The base resolution is $24 \times 80 \times 32$, and we use 4 levels of refinement, fully refining the region around the loop, resulting in an $x-y$ plane resolution of 31.2 km per cell, or $0.02R$, where $R$ is the radius of the loop. The fully-refined resolution in the $z$ direction is 234 km per cell. Comparison with simulations with one more level of refinement shows no important differences in the dynamics, even though the KHI instability (and later turbulence) changes quantitatively.
  
\section{Results and discussion}

\begin{figure*}
    \centering
        \includegraphics[width=0.75\textwidth]{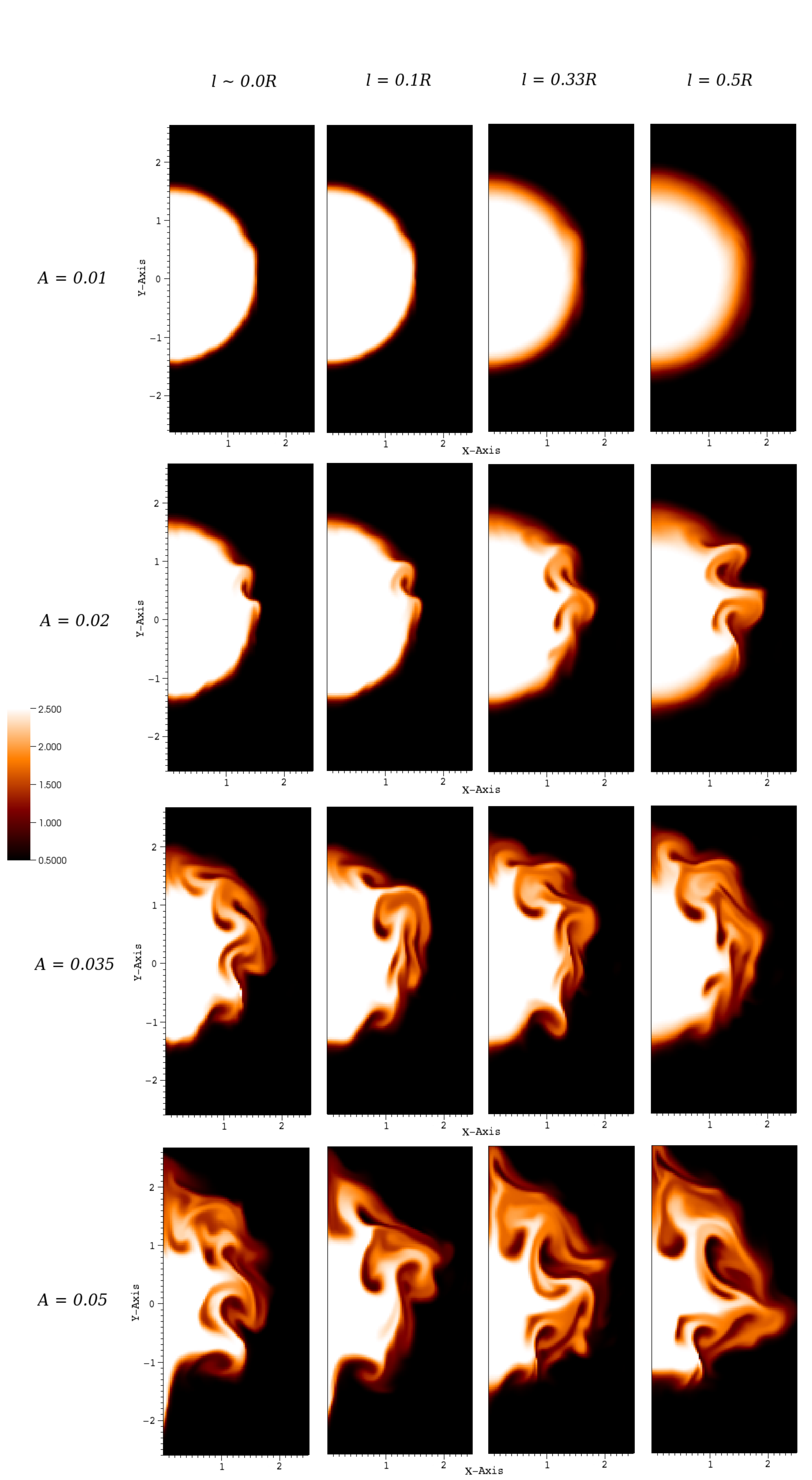}  
        \caption{Plots showing the density in the anti-node cross-section of the loop ($z = 0$) at $t = 540$ s for the different parameters used ($A=0.005$ not shown for brevity, looking almost identical to $A=0.01$). Different columns have different initial inhomogeneous layer widths, shown at the top, and the different rows are for different amplitudes, shown in the left margin. Axis units are in Mm and density in units of $10^{-12}\ \mathrm{kg\ m^{-3}}$. }
        \label{dens540}
 \end{figure*}

\begin{figure*}
    \centering
        \includegraphics[width=1.0\textwidth]{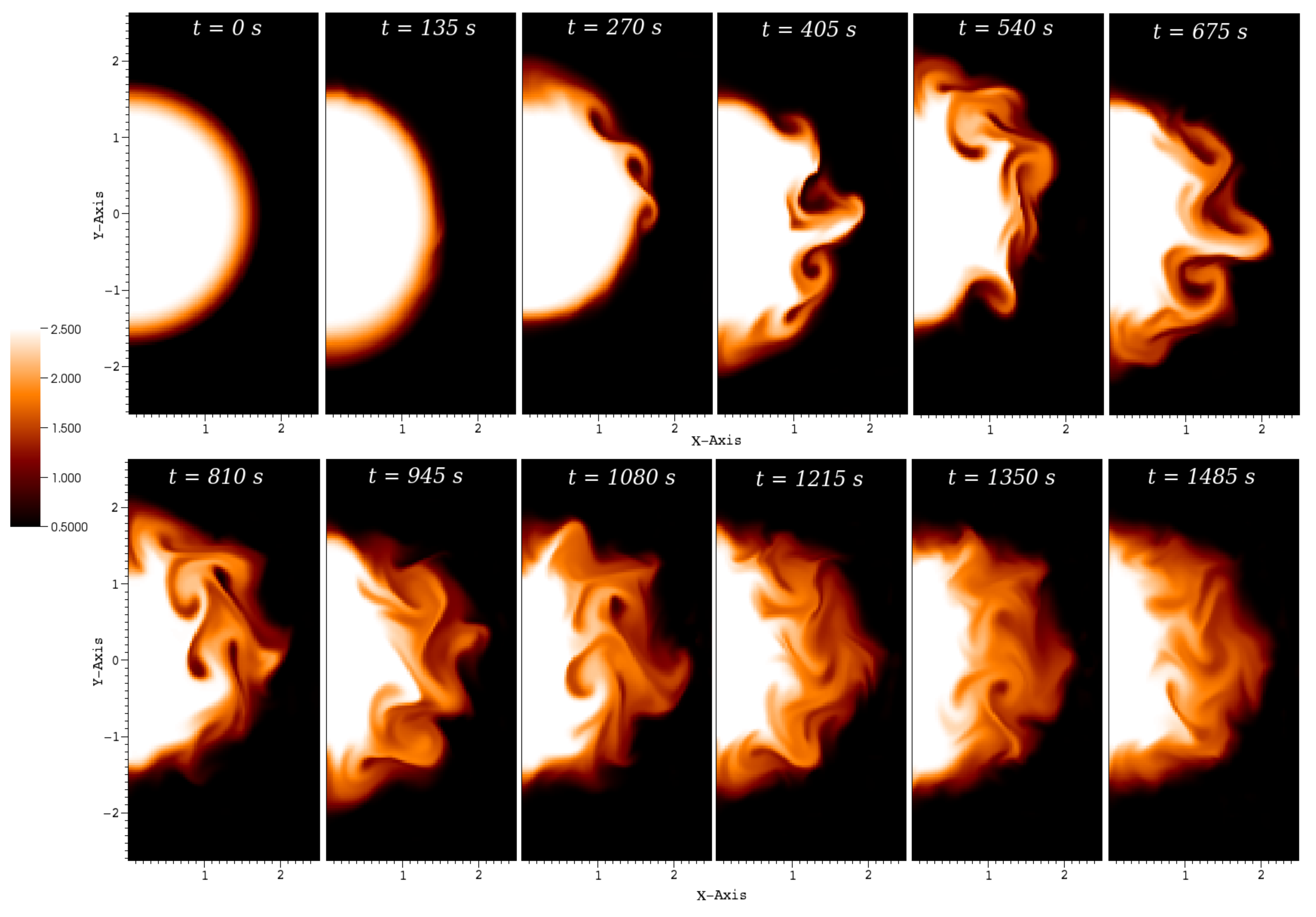}  
        \caption{Plots of density in the anti-node cross-section of the loop ($z = 0$) at different times (in steps of approximately one half-period, shown in the upper part of each plot) for $A = 0.035$ and $l = 0.33R$. Axis units are in Mm and density in units of $10^{-12}\ \mathrm{kg\ m^{-3}}$. }
        \label{dens0.035}
 \end{figure*}

\begin{figure*}
    \centering
        \includegraphics[width=1.0\textwidth]{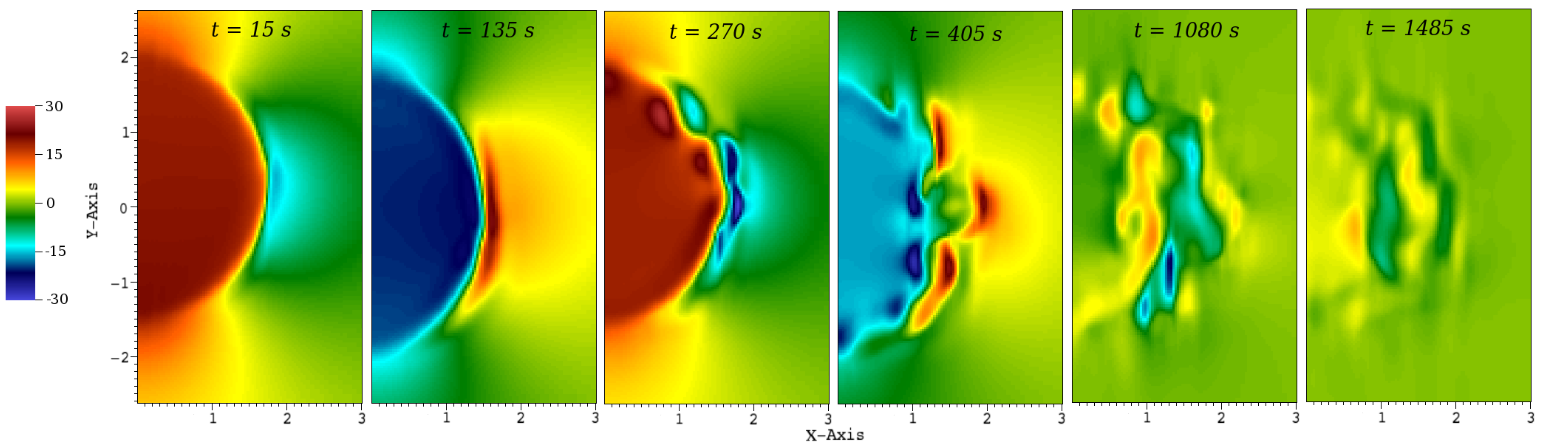}  
        \caption{Plots of the $y$-component of the velocity ($v_y$) at different times (selected times corresponding to plots in Figure~\ref{dens0.035}, shown in the upper part of each plot), for $A = 0.035$ and $l = 0.33R$. Axis units are in Mm and velocity in units of km s$^{-1}$.}
        \label{vy0.035}
 \end{figure*}

\begin{figure*}
    \centering
       \begin{tabular}{@{}cc@{}}
        \includegraphics[width=0.5\textwidth]{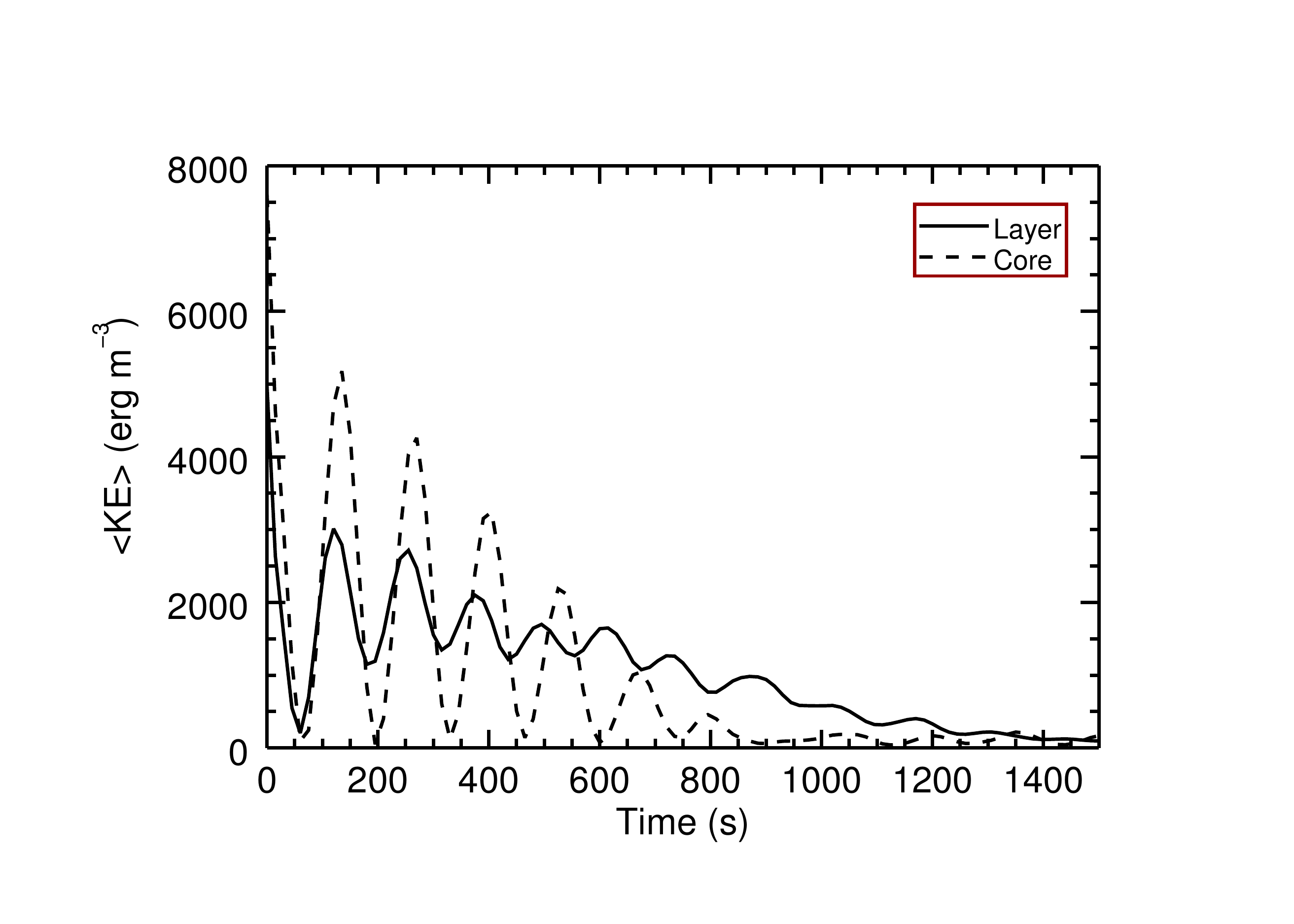}  
        \includegraphics[width=0.5\textwidth]{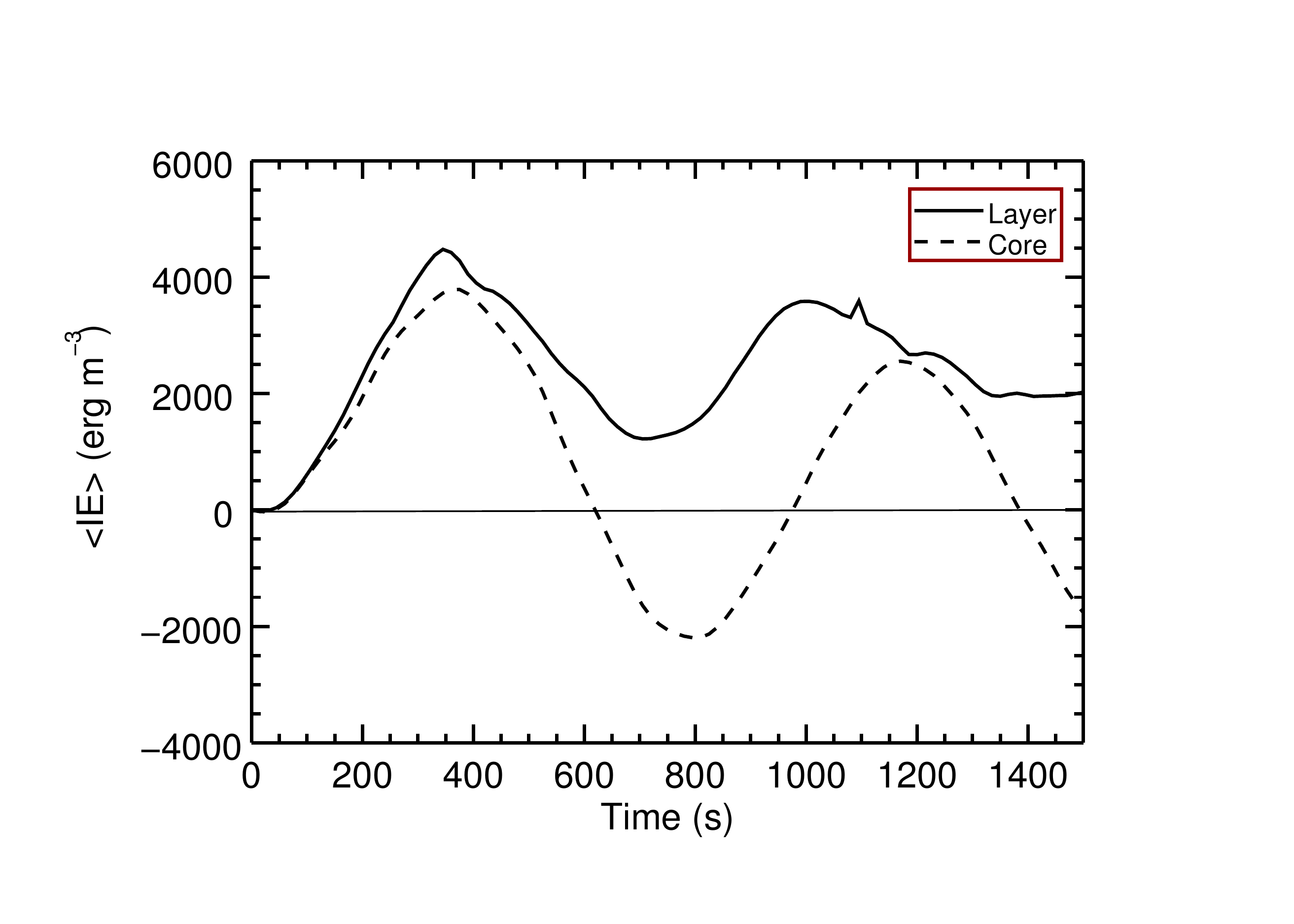} \\
        \includegraphics[width=0.5\textwidth]{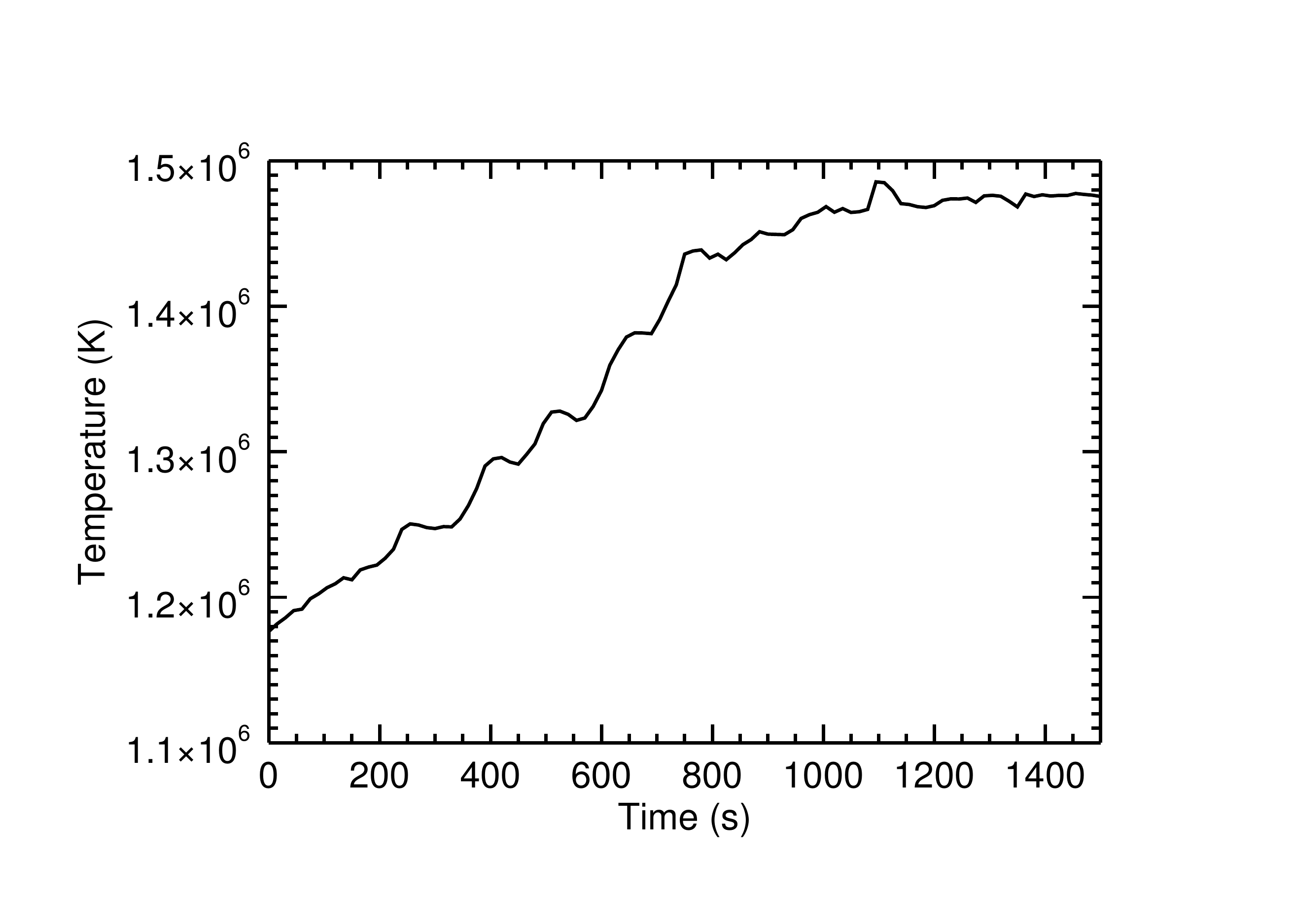} 
        \includegraphics[width=0.5\textwidth]{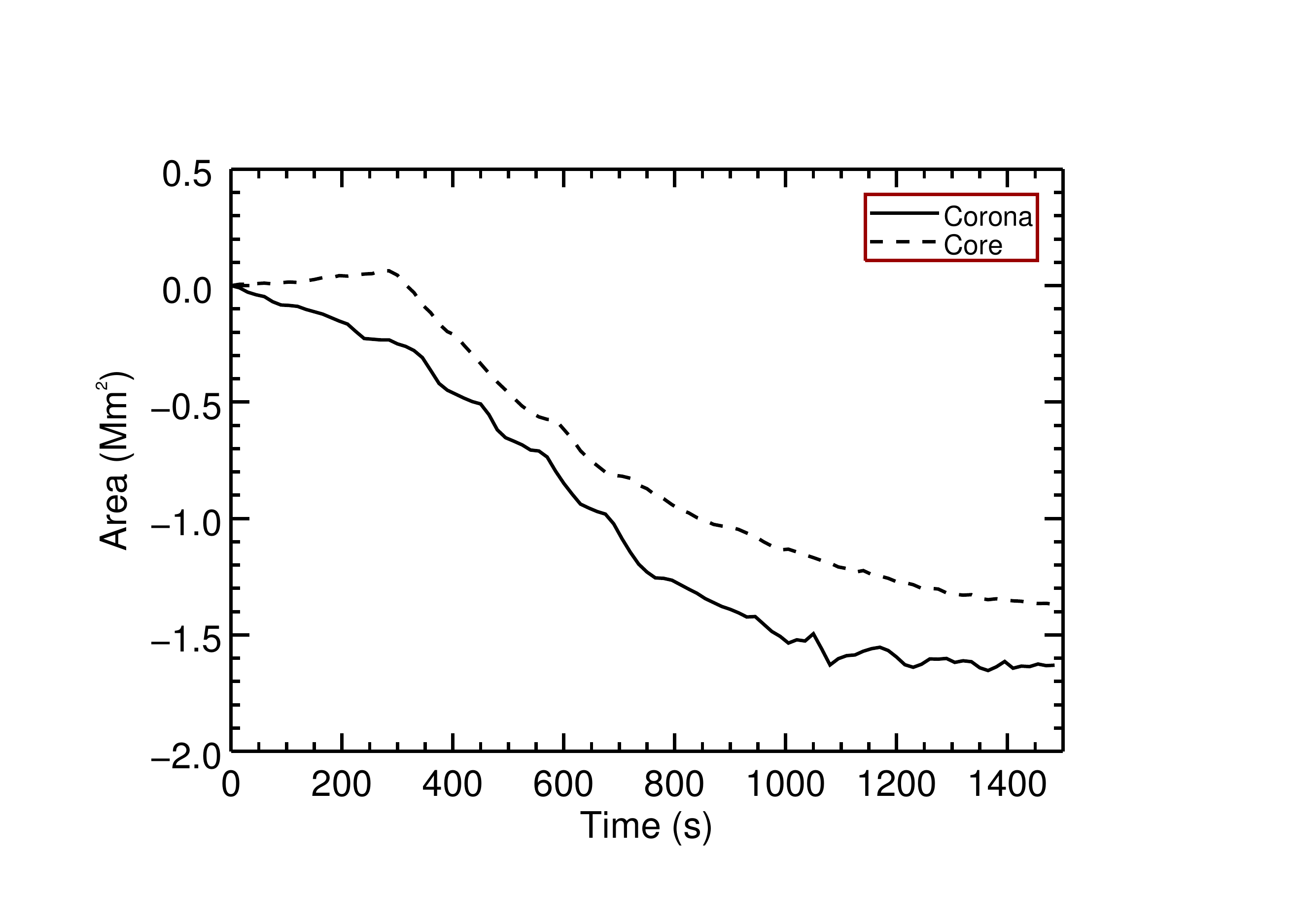} 
       \end{tabular}  
        \caption{Plots showing the time evolution of different quantities in the anti-node cross-section of the loop ($z = 0$) with $A = 0.035$ and $l=0.33R$. \textit{Top-left}: Average kinetic energy (KE) density in the core and layer region of the loop. \textit{Top-right}: Average internal energy density in the core and layer regions of the loop. \textit{Bottom-left}: Average temperature of the loop (average over both the core and layer). \textit{Bottom-right}: Change in the area relative to the initial value, for the core region of the loop and the numerical domain outside the loop (outside both core and layer). }
        \label{0.035KE}
 \end{figure*}

\begin{figure*}
    \centering
       \begin{tabular}{@{}cc@{}}
        \includegraphics[width=0.5\textwidth]{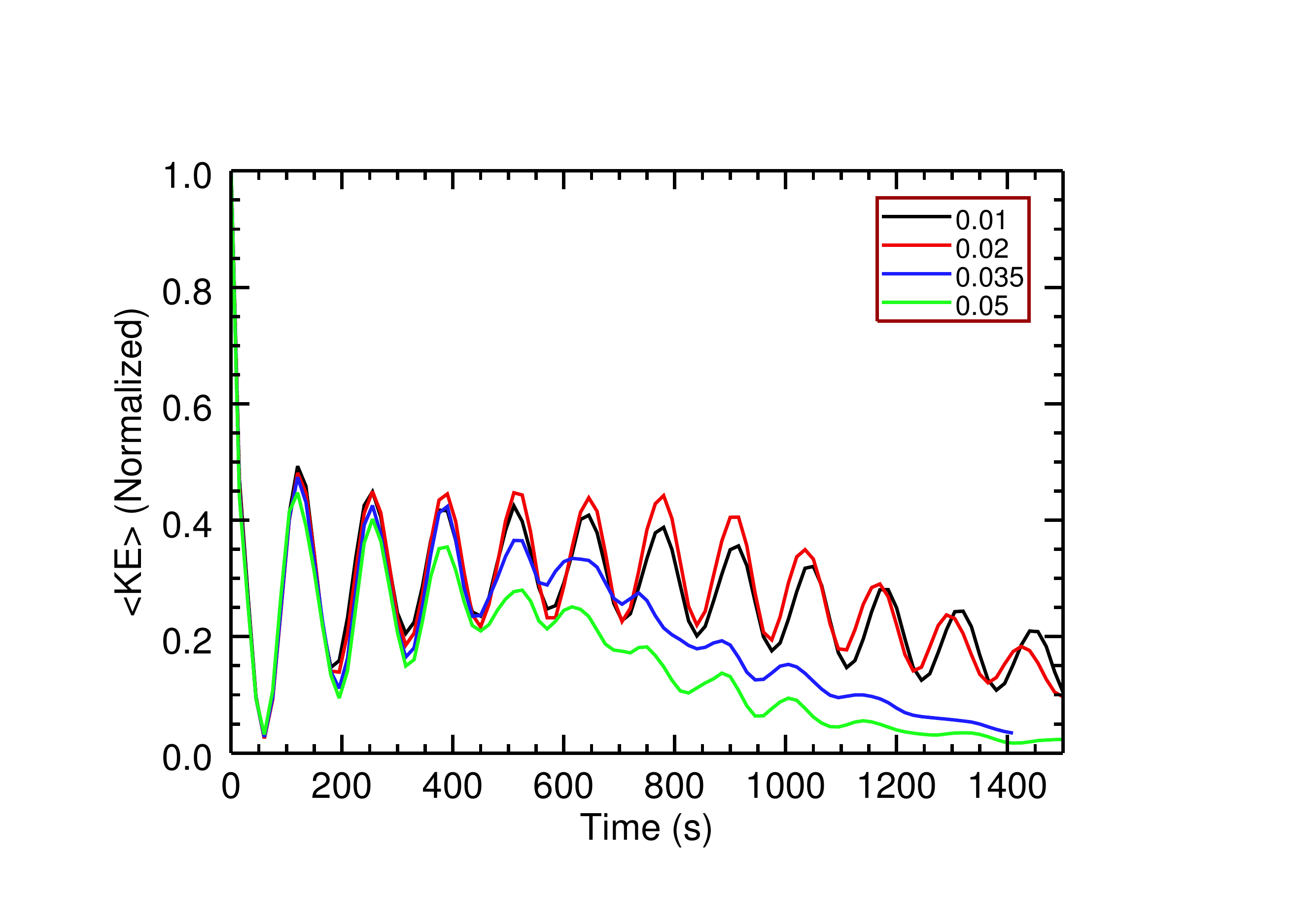}  
        \includegraphics[width=0.5\textwidth]{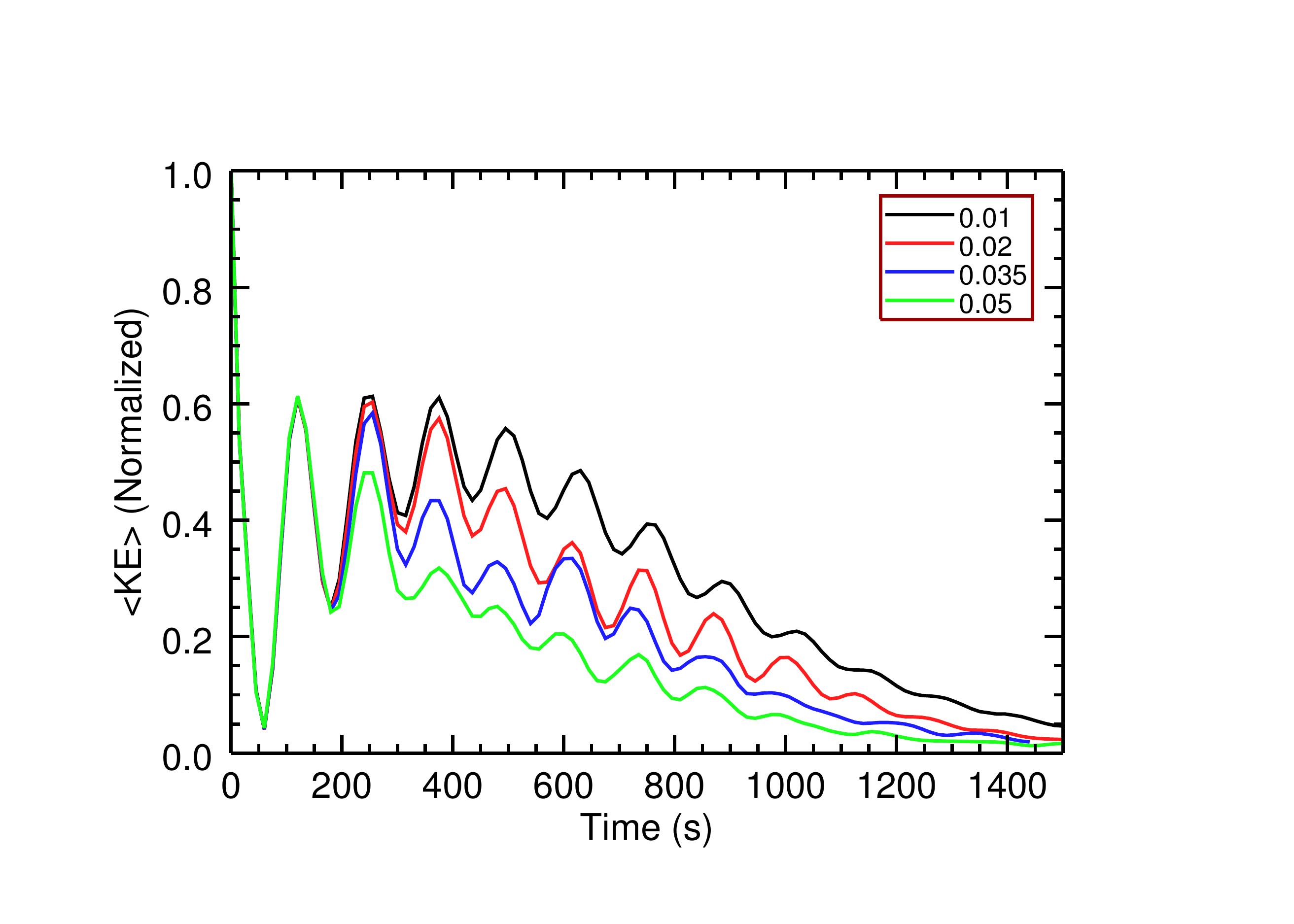} 
       \end{tabular}  
        \caption{Plots of the average kinetic energy density over time in the inhomogeneous layer of the anti-node cross-section of the loop ($z = 0$). \textit{Left}: $l = 0.1R$. \textit{Right}: $l = 0.5R$. The different lines are for different initial amplitudes, from $0.01$ to $0.05$.}
        \label{KE}
 \end{figure*}

\begin{figure*}
    \centering
       \begin{tabular}{@{}cc@{}}
        \includegraphics[width=0.5\textwidth]{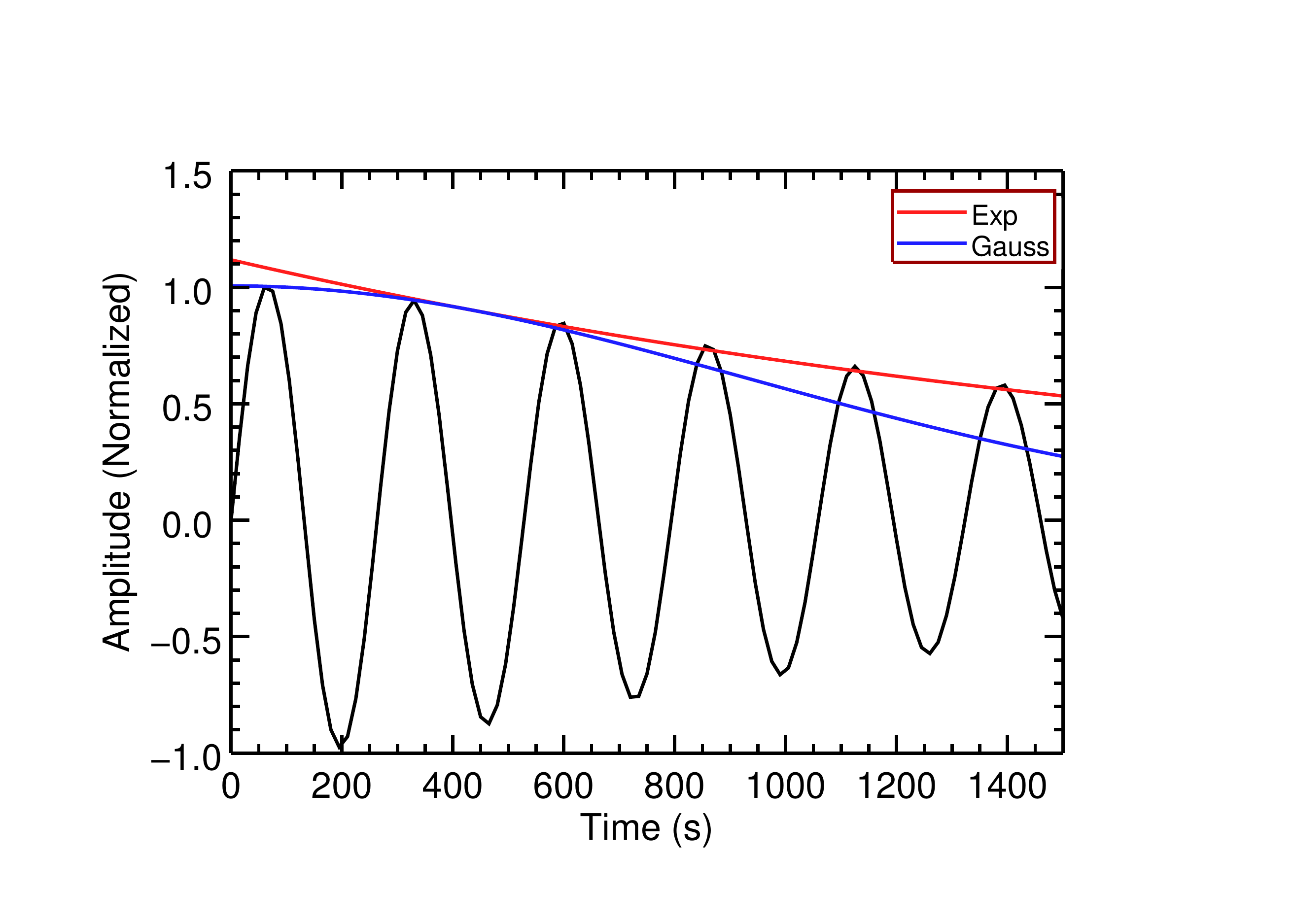}  
        \includegraphics[width=0.5\textwidth]{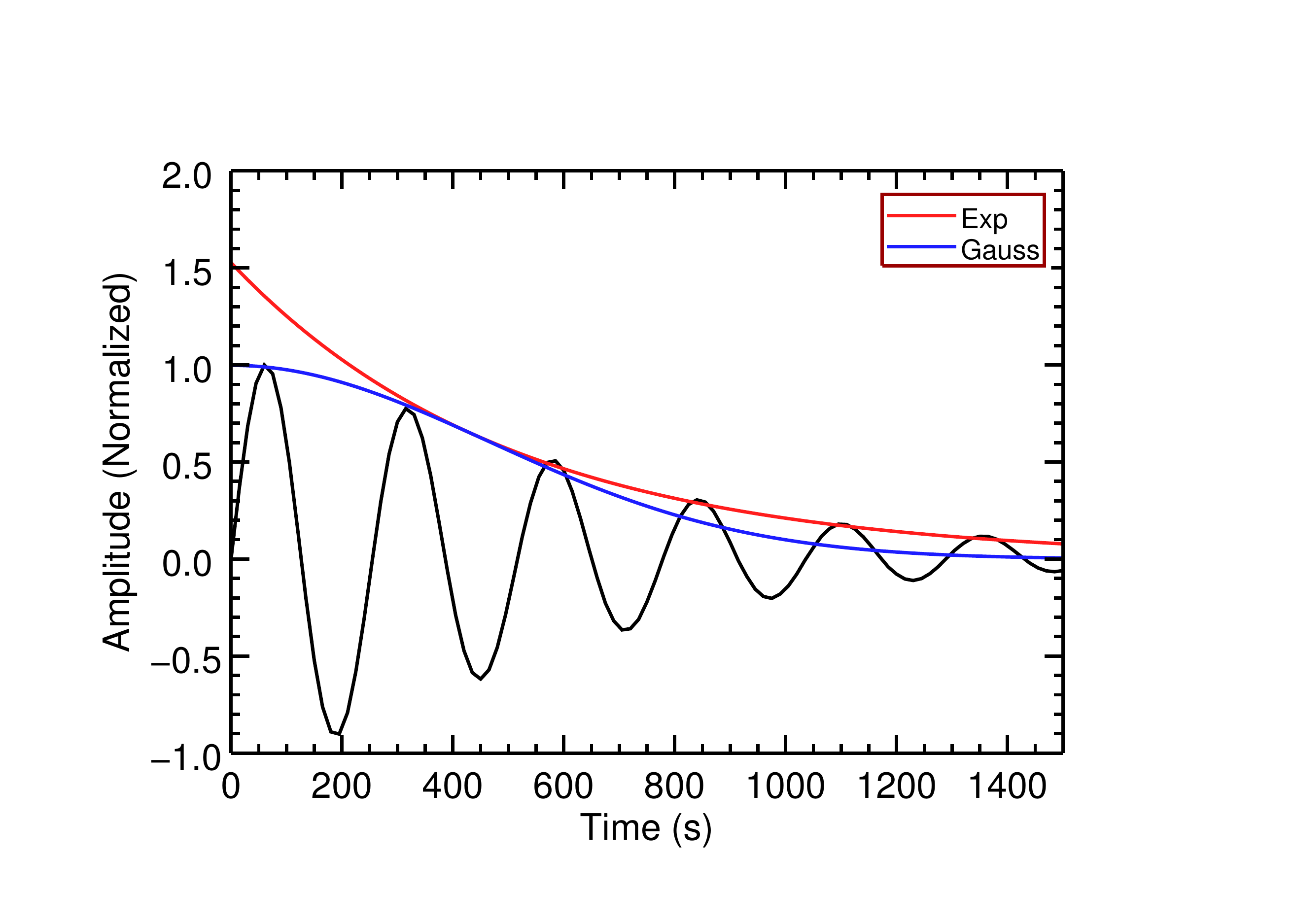} \\
        \includegraphics[width=0.5\textwidth]{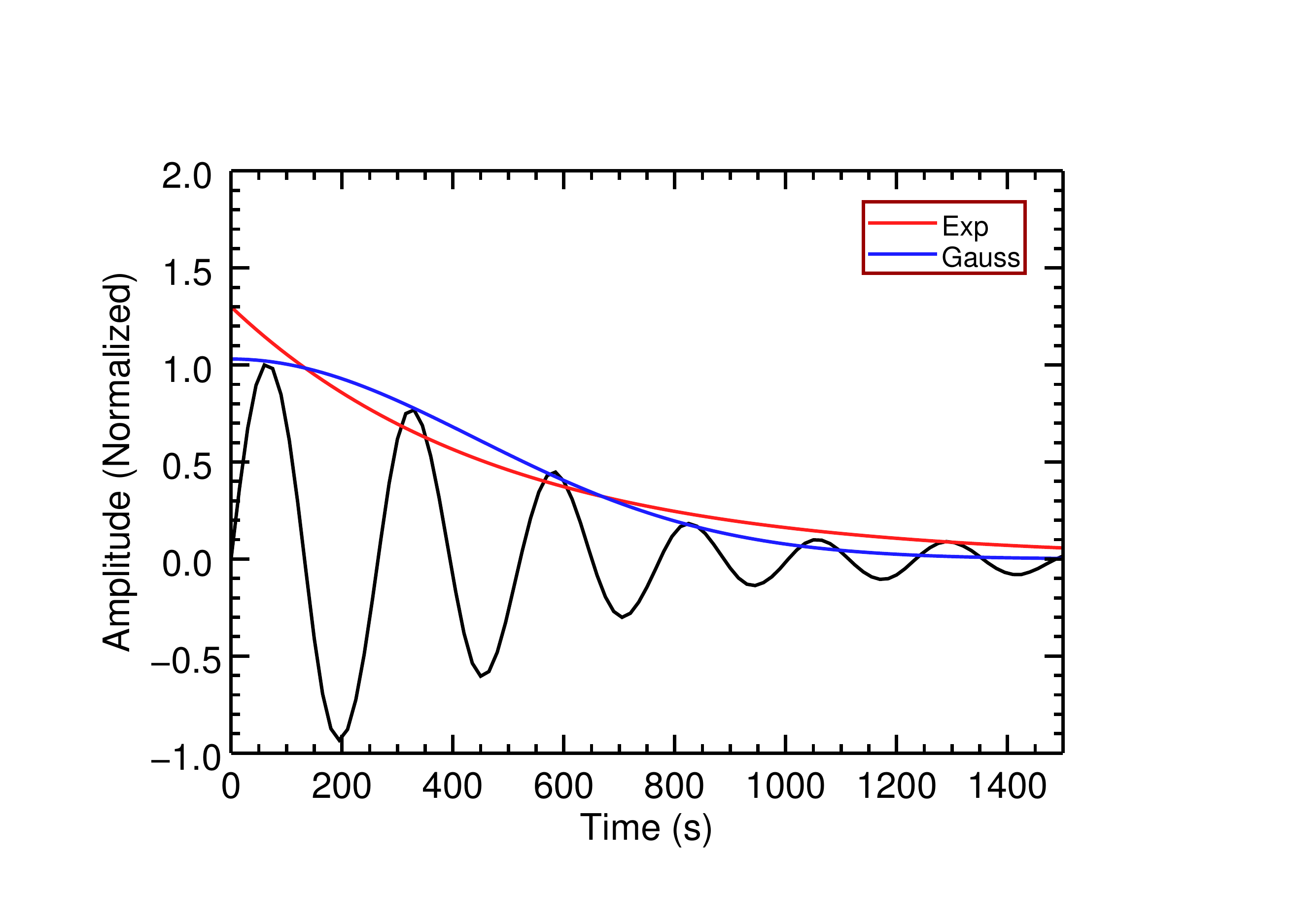} 
        \includegraphics[width=0.5\textwidth]{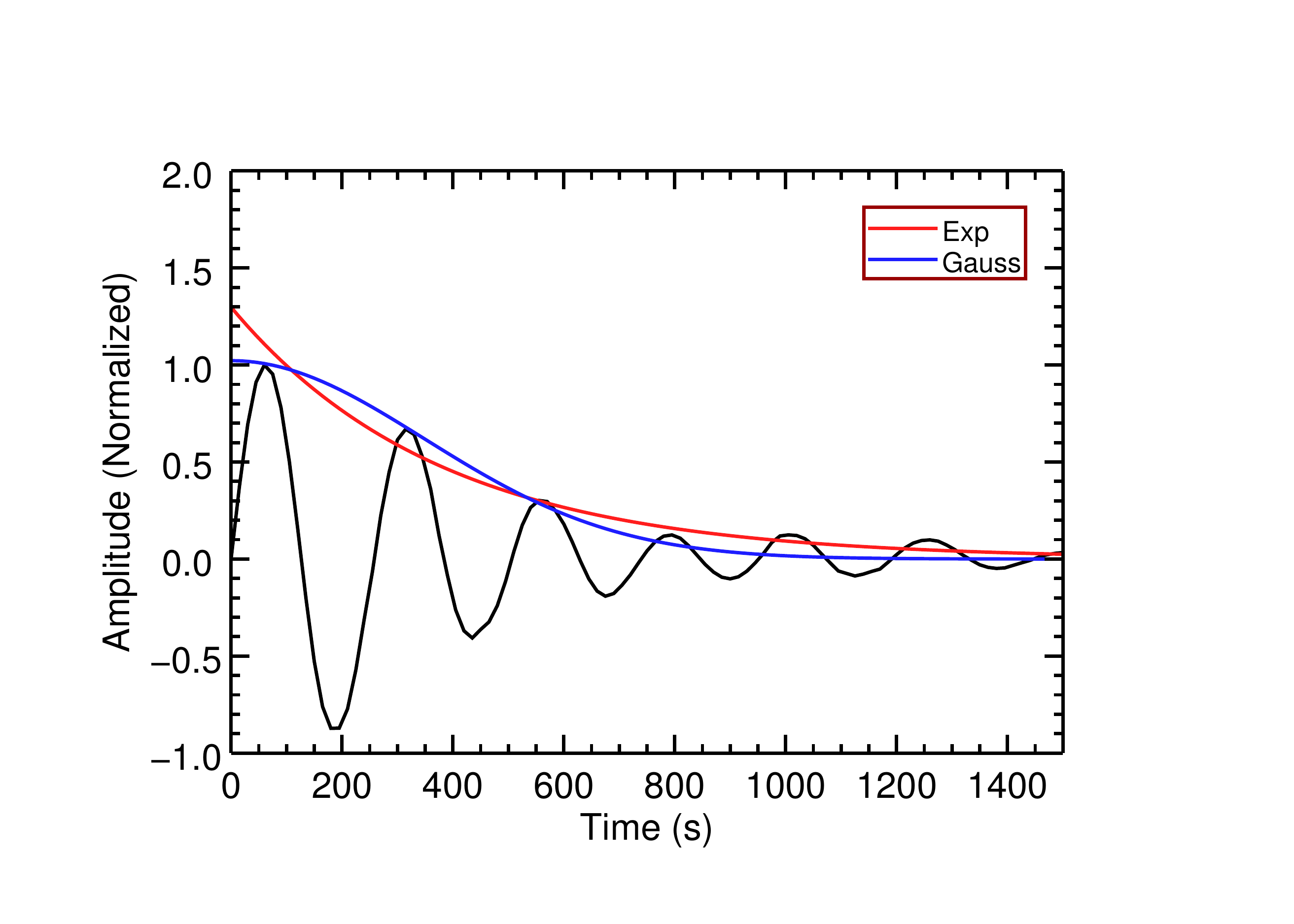} 
       \end{tabular}  
        \caption{Plots of the displacement amplitude over time in the loop anti-node ($z=0$). The values are normalized to the initial displacement. For the top plots ($A = 0.01$, \textit{Left}: $l=0.1R$, \textit{Right}: $l=0.5R$), the Gaussian and exponential profiles from Eq.~\ref{non-ossc} obtained by fitting Eq.~\ref{fit} are plotted. For the bottom plots ($A = 0.05$, \textit{Left}: $l=0.1R$, \textit{Right}: $l=0.5R$), the Gaussian of Eq.~\ref{non-ossc} is fitted for the whole data and plotted, as well as the exponential resulting from fitting Eq.~\ref{expon}.}
        \label{fitpic}
 \end{figure*}

\begin{figure}
    \centering
        \includegraphics[width=0.5\textwidth]{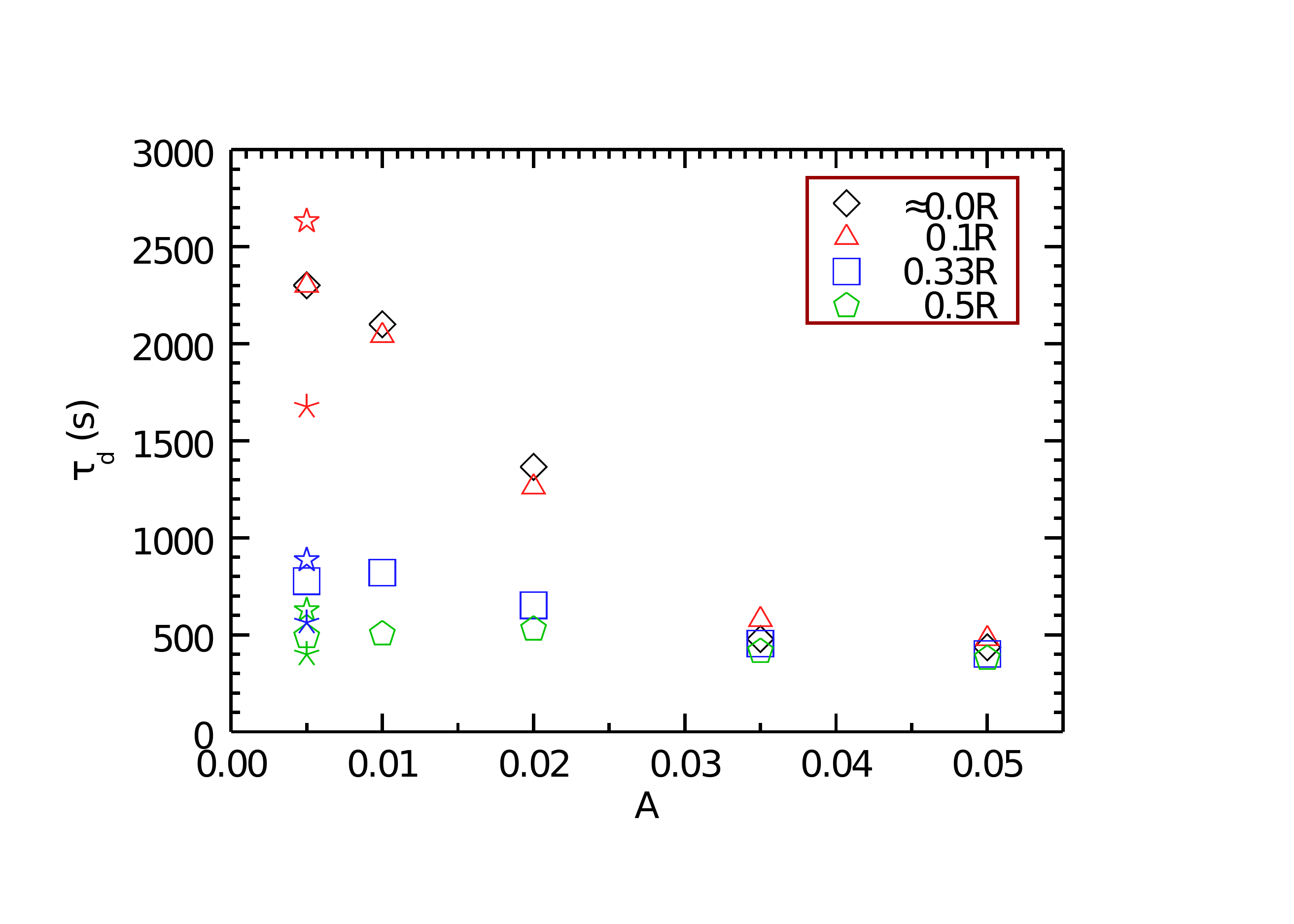}
        \caption{Plot showing the exponential damping times for the different initial amplitudes, ($A$), and inhomogeneous layer widths, represented by diamond ($\diamond$), triangle ($\triangle$), square ($\sq$), and pentagon ($\pentagon$), in increasing order. The stars ($\star$) of the same colour as the symbols, represent the theoretically predicted damping time for the specific layer width and sinusoidal layer density profile. Analogously, the asterisks ($\APLstar$) represent the predicted damping for a linear layer density profile \citep{2002A&A...394L..39G}.}
        \label{damping}
 \end{figure}

\begin{figure}
    \centering
         \includegraphics[width=0.5\textwidth]{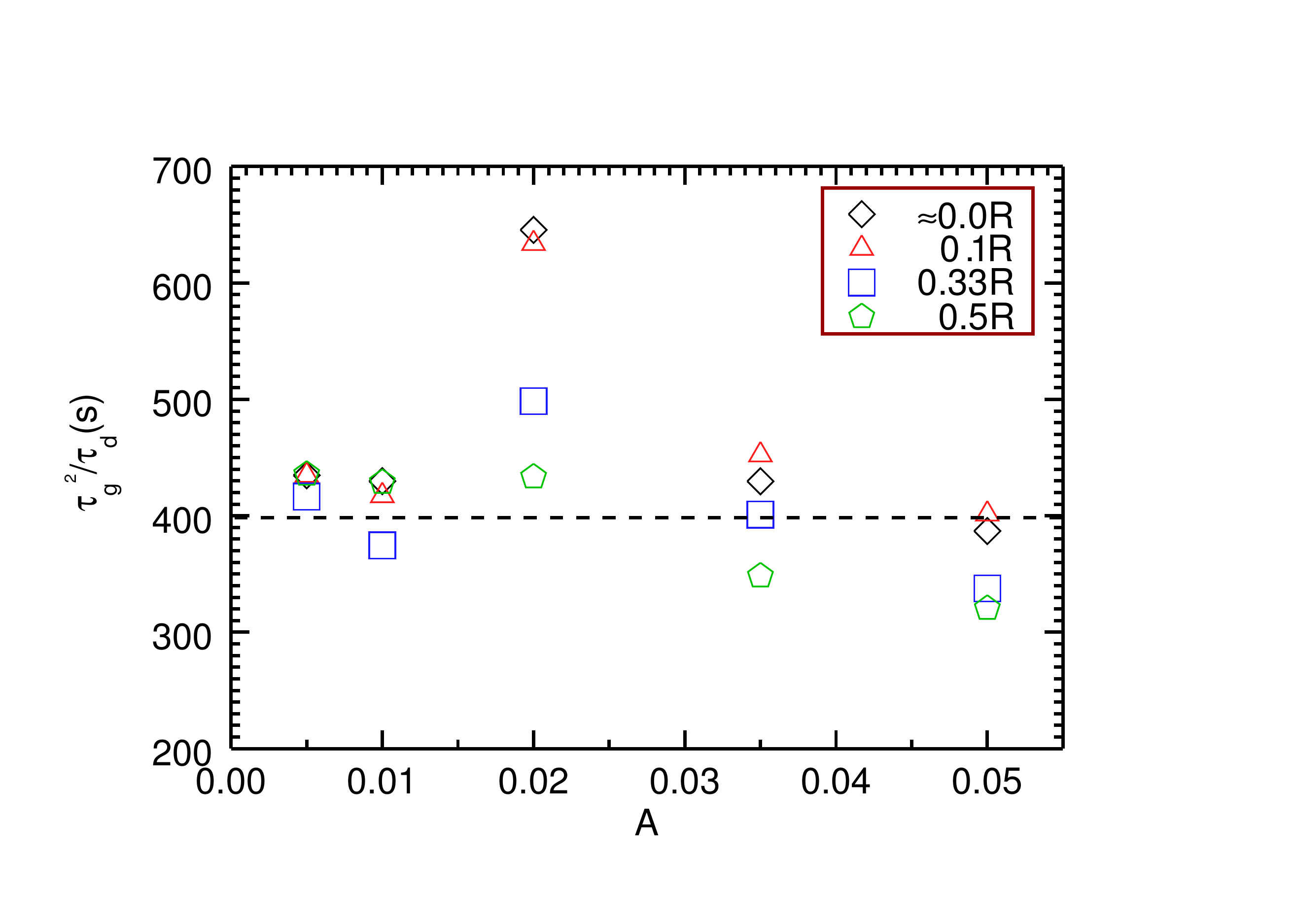}
    \caption{Plot showing the switch time $t_\mathrm{s}$ (Eq.~\ref{tied}) between the Gaussian and exponential damping, obtained through fitting as described in the article body. The symbols used for different layer widths are the same as in Figure~\ref{damping}. The dashed line at 400 s is the theoretical value of $t_\mathrm{s}$, for our parameters.}
    \label{tgtd}
 \end{figure}

We ran a series of simulations which explored the parameter space, varying the initial velocity perturbation and thickness of the inhomogeneous layer of the loop. The chosen values are the following: $A = \{0.005,0.01,0.02,0.035,0.05\}$ and $l = \{{\approx}0.0R,0.1R,0.33R,0.5R\}$, for a total of 20 different runs (Figure~\ref{dens540}). 
The first value for $l$ denotes an initially step profile which, after the start of the simulation, evolves into an inhomogeneous layer due to numerical diffusion. This layer is thus the thinnest possible layer in our simulation, corresponding to approximately $0.08R$. The initial displacements of the loop are, in increasing order with the amplitude, $ \{0.079R, 0.16R,0.32R,0.55R,0.78R\}$. Obtained oscillation periods are discussed later in the text. The nonlinearity parameter in the case of kink oscillations of a flux tube is  given approximately by \citet{2014SoPh..289.1999R}:
\begin{equation}
 \nu \simeq \frac{A L}{R}
.\end{equation}
The oscillations are weakly nonlinear for $\nu \ll 1$ and strongly nonlinear for $\nu \geq 1$. For the amplitudes chosen, the nonlinearity parameter varies between 0.4 and 4. Thus, all the simulations represent nonlinear and strongly nonlinear regimes of oscillations. As the simulations with $A = 0.005$ and $A = 0.01$ show almost similar evolution, lower amplitudes were not essential for this study. In Figure~\ref{dens0.035}, the evolution of the oscillation in the anti-node cross-section can be followed, for a specific set of parameters.
In this sequence, we can identify the principal signatures of nonlinear kink oscillations previously described in detail in numerical simulations by \citet{2008ApJ...687L.115T,2014ApJ...787L..22A,2015ApJ...809...72A,2015A&A...582A.117M}. These works however, lack a description of one aspect which plays a role in the nonlinearly enhanced damping of the oscillations, namely the resonance between the kink and $m \geq 2$ or fluting modes, which extracts energy from the kink mode, leading to damping of the transverse oscillation. This effect was studied analytically by \citet{2010PhPl...17h2108R}, and is present only in the absence of longitudinal inhomogeneities (e.g. gravitational stratification), which destroys the resonance between the modes \citep{2014SoPh..289.1999R}. This effect can be seen clearly in the second snapshot of Figure~\ref{dens0.035} ($t=135$ s, approximately half-period), before the development of the Kelvin-Helmholtz instability (KHI),  as a deformation of the initially circular cross-section in a manner resembling the $m=2$ fluting mode. 
There are three main mechanisms acting to damp the transverse oscillation (aside from a very small numerical dissipation): resonant absorption or mode conversion, the previously mentioned $m \geq 2$ resonance, and the development of the KHI (in the case shown in Figure~\ref{dens0.035}, in less than a period). These mechanisms act concomitantly and are coupled, resulting in the significance and magnitude of each mechanism varying in time, which is difficult to estimate. The kink mode presents a velocity shear near the lateral boundary of the loop, which is prone to the KHI. Resonant absorption, acting immediately after $t=0$, further enhances the velocity shear, thus shortening the time after which the instability sets in.  Note that, even in the absence of an inhomogeneous layer (which, due to numerical diffusivity, is not possible in our simulations), the KHI can still develop. In this sense, in the nonlinear regime, transverse oscillations will always be damped.\\
As previously stated, estimating the relative importance of the interacting damping mechanisms is not an easy task. By looking at the transverse velocity ($v_y$), in Figure~\ref{vy0.035}, we can appreciate the interaction of two mechanisms, resonant absorption and KHI.
Initially, the analytical kink eigenfunction for velocity develops after the perturbation at $t=0$, representing a uniform speed within the boundaries of the loop and a dipolar field around it. As noted earlier, a velocity shear is present in this configuration, which, as shown
in the $t=135 s$ plot, is later further enhanced by resonant absorption. Later on, the resonant layer is disrupted, its KE contributing to the development of the KHI. Due to the robust nature of resonant absorption \citep{2008ApJ...679.1611T}, it is still present in patches, that is, on the surface of the roll-ups, but ceases to exist in the form which was present at $t=135 s$. It is difficult to estimate the relative effectiveness and the contribution to the damping of the patchy resonant absorption compared to the pre-disruption phase. Ultimately, the roll-ups break up, resulting in a turbulent inhomogeneous layer, where most of the remaining wave energy is situated. The impact of the $m$-mode resonance on the damping time is not clear.  \citet{2015A&A...582A.117M} studied nonlinear fundamental standing kink oscillations in a stratified atmosphere, in which the resonance is still 
noticeable, despite the longitudinal inhomogeneity. In Figure 2 of \citet{2010PhPl...17h2108R} the enhanced damping as a function of a nonlinearity parameter N is shown, which is roughly equal to the driving amplitude divided by the width of the inhomogeneous layer in units of $R$. For our parameters, this value is at most 0.5. This value corresponds to a very small effect on the oscillation amplitude. In this sense, we can be somewhat confident that this damping mechanism does not have a substantial impact on the resulting damping times, and in what follows we focus on the other damping mechanisms.\\
We studied the energy evolution in the anti-node cross-section of the loop for the different parameters. For the case in Figure~\ref{dens0.035}, the average values of relevant variables are plotted as a function of simulation time in Figure~\ref{0.035KE}.
Note that, in the anti-node of the fundamental kink, all of the kink mode's energy is in the form of KE. The oscillations present in this plot, with the double period of the kink mode, represent the exchange between KE and magnetic energy (ME), analogous to the exchange between kinetic and elastic potential energy in a classical harmonic oscillator (if we account only for the perturbation to the ME). To proceed further, we make the distinction between  the core of the loop, defined as the region of the cross-section where $\rho \geq 0.98 \rho_i$ and the ME is negligible (see \citet{2014ApJ...788....9G} for example), and the inhomogeneous layer, defined as the region of the cross-section where $0.98\rho_i \geq \rho \geq 1.1\rho_e$. Inspecting the figures, we observe that the average KE density in the core of the loop has a decreasing peak energy. At the same time the energy contained in the inhomogeneous layer is undergoing a different evolution: the peak values are decreasing, but its minimal values show a strong initial increase, which can be attributed to the increase in the energy of localized Alfvén waves, through resonant absorption (see, e.g. \citet{1991PhRvL..66.2871P}). Simultaneously, the KE in the layer is dissipated, resulting in an increase of internal energy (IE). The average IE density shows a long-period oscillation, triggered by the perturbation at $t=0$, corresponding to a longitudinal slow mode. The slow mode is present both in the core and the layer, however the background trend of average IE density shows a steady increase in the layer, in accordance with the KE loss. The IE density plot shows the average perturbation to its equilibrium value over time, and the rise due to the dissipation of the KE represents a $\approx0.36\% $ increase in IE density. This cannot account for the apparent heating of the loop (core with layer), from an average temperature of $1.17\ \mathrm{MK}$ to $1.47\ \mathrm{MK}$: this would require around 50 times more energy converted into IE. Inspecting the change in the area of the core and the surrounding corona (outside the loop, area of the numerical domain), we notice that both shrink, resulting in an enlargement of the inhomogeneous layer. However, we also notice that the core area diminishes to a lesser extent  than the corona. This implies that the higher average temperature of the loop is due to the enlarging inhomogeneous layer, mixing with the hot and rarefied plasma surrounding the loop, driven by the KHI. This effect is present only because of the chosen initial conditions, that is, with a corona five times hotter surrounding the loop. Note that the total IE in the layer (not shown here) presents a considerable increase ($\approx 150\%$), but this is almost entirely (i.e. except the minute increase converted from KE) a result of the enlargement of the layer, caused by the KHI. \\\
In the plots of Figure~\ref{KE} the evolution of the average KE density in the inhomogeneous layer for different initial amplitudes and layer thickness can be compared.
These plots are normalized to the initial energy density contained in the layer, (i.e. of the initial perturbation), so that a comparison for different amplitudes is straightforward. We note that the relative increase of KE density in the layer is higher for smaller amplitudes, and it requires more time to dissipate. This demonstrates that the presence of KHI in the higher amplitude oscillations acts to enhance the conversion of KE to IE. This effect is more pronounced for wider initial inhomogeneous layers. \\
In the following, we analyze the evolution of the kink oscillations triggered by the perturbation and their associated damping. For this, we track the centre of mass for the loop in the fundamental kink anti-node cross-section where the displacement is maximum. The damping profile of  impulsively triggered standing kink modes in a flux tube is not purely exponential \citep{2013A&A...551A..39H,2013A&A...551A..40P,2016A&A...589A.136P}. Instead, initially the damping is approximated by a Gaussian profile, and after a time $t_\mathrm{s}$ by an exponential profile. This profile has the form: 
 \begin{equation}
 A(t) = \left\{
  \begin{array}{lr}
    A_0\ \mathrm{exp} \left(-\frac{t^2}{2 \tau_\mathrm{g}^2}\right) &  t \leq t_\mathrm{s} \\
    A(t_\mathrm{s})\ \mathrm{exp} \left(-\frac{t-t_\mathrm{s}}{\tau_\mathrm{d}}\right) &  t > t_\mathrm{s}
  \end{array}
 \right. 
 \label{non-ossc}
 ,\end{equation}
and the switch time $t_\mathrm{s}$ is defined as:
 \begin{equation}
  t_\mathrm{s} = \frac{\tau_\mathrm{g}^2}{\tau_\mathrm{d}} = \frac{\rho_0/\rho_e + 1}{\rho_0/\rho_e - 1}P
  \label{tied}
 .\end{equation}
Thus, by fitting observed standing kink oscillations of coronal loops to a function 
 \begin{equation}
 A(t)\  \mathrm{sin}(\omega t + \phi)
 \label{fit}
 ,\end{equation}
one could seismologically estimate the density ratio of the loop and inhomogeneous layer thickness \citep{2016A&A...589A.136P}. We fit Eq.~\ref{fit}, using \texttt{mpfitfun.pro} \citep{2009ASPC..411..251M}, to the simulated damping profiles to obtain four parameters, $A_0, \tau_\mathrm{g}, \tau_\mathrm{d}$, and $\omega = \frac{2\pi}{P}$. Note that $t_s$ is a constrained parameter (Eq.~\ref{tied}) using the `\texttt{.TIED}' entry of the \texttt{PARINFO} structure, and $\phi$ is zero as it results from the initial condition. We find best-fittings for the period between 253.5 and $ 265.1\ \mathrm{s}$; thicker inhomogeneous layers resulting in the shorter periods. The period for the simulation closest to the linear case ($A = 0.005,\ l \approx 0.0R$) is 265.1 s. The theoretical fundamental kink period for a step-density flux tube with the same parameters is $263.6\ \mathrm{s}$ \citep{1983SoPh...88..179E}. Some results from fitting to the simulated damping profiles are shown in Figure~\ref{fitpic}.
The amplitudes were normalized to the value of the maximal initial displacements. Note that for the low-amplitude case ($A = 0.01$, upper plots), the fit is much more accurate (lower $\chi^2$) than for the high-amplitude counterpart ($A = 0.05$, lower plots). Furthermore, we find that, in the runs which develop KHI ($A \geq 0.02$), the last periods of oscillation are practically undamped. This can also be seen in Figure~\ref{0.035KE} (top left), in the core average KE, which remains practically constant in maximal value after $\approx 1000$ s. We interpret this as a consequence of the KHI: some of the kinetic energy percolates into the core region. Although a small effect, this unfortunately makes the fitting of Eq.~\ref{fit} meaningless (i.e. a Gaussian profile will always fit better to the oscillations, for any value of $t_s$). Overcoming this problem by truncating the dataset where the constant KE period begins (e.g. 800 s) is unreliable as $t_s$ varies greatly even for small changes in the truncation. Thus, great care should be taken when fitting Eq.~\ref{fit} to observed oscillations for seismology. The oscillations resulting from $A \geq 0.02$ are best described by a Gaussian profile. Therefore, we encourage the fitting of Gaussian damping profiles alongside the traditional exponential damping profile in future studies of transverse coronal loop oscillations. For our results to be readily comparable to available studies, we quantify the exponential damping time, and to obtain $t_s$ for the high amplitude oscillations, we fit an exponentially damped sine;
\begin{equation}
A(t) = A_0\ \mathrm{exp}\left(-\frac{t}{\tau_\mathrm{d}}\right)\ \mathrm{sin}(\omega t)
\label{expon}
,\end{equation}
for the whole dataset. The exponential damping times obtained in this way are plotted in Figure~\ref{damping}. 
Note that for `well-behaving' amplitude profiles (for $A \leq 0.01$) we show $\tau_\mathrm{d}$ obtained from Eq.~\ref{fit}. We also plotted the theoretical damping times due to resonant absorption for different inhomogeneous layer widths, for both linear and sinusoidal density profiles in the layer \citep{2002ApJ...577..475R,2002A&A...394L..39G}. We find that, for the low amplitude cases ($A \leq 0.01$) the simulated damping times are in between the predicted ones: the density profile in the inhomogeneous layer can greatly influence the resulting theoretical damping time, especially for thin layers \citep{2013ApJ...777..158S}. We would like to highlight three important characteristics of the nonlinear damping of kink oscillations from Figure~\ref{damping}: 
\begin{itemize}
\item[$\bullet$] The damping time is no longer generally independent of amplitude, as in the linear case. There is a strong damping, even with initially thin inhomogeneous layers of the kink oscillations, for high enough amplitudes. In these cases, the damping time can be less than a third of the theoretically predicted one. Furthermore, the damping time appears to be saturated for amplitudes larger than some threshold value.
\item[$\bullet$] For thick enough inhomogeneous layers (in our case, for $l = 0.5R$), the damping time is independent of the amplitude, and thus coincidentally well approximated by the theory.
\item[$\bullet$] In the saturated high amplitude regime (here for $A \geq 0.035$), the damping is only weakly dependent on the initial thickness of the inhomogeneous layer, and can thus not be used for seismological purposes.
\end{itemize}
Having in mind the aforementioned characteristics, we can state that high amplitude oscillations of coronal loops (as induced by a flare or Low Coronal Eruption (LCE), see \citet{2015A&A...577A...4Z}) are constrained to damp with specific damping times, only weakly dependent on the amplitude and the initial inhomogeneous layer width. This behaviour is associated with the presence of the KHI in the high amplitude oscillations. Thus, the growth rate of the KHI determines the switch between the linearly well-approximated damping and high amplitude damping regimes. For an analytical estimate of the growth rate of KHI for standing transverse oscillations, see \citet{2015ApJ...813..123Z}. In our case, this switch is around $A = 0.02$, which can be seen as a transitional regime: the KHI is developing but not fully, and is not disrupting the classical $m = 1$ resonant absorption (as at $t= 135$ s in Figure~\ref{vy0.035}). \\ 
For our initial conditions and measured oscillation periods, the time of transition between Gaussian and exponential damping profiles (Eq.~\ref{tied}) is at $t_\mathrm{s} \approx 400$ s. In Figure~\ref{tgtd} we plotted the values for $t_\mathrm{s}$ obtained through the fitted Gaussian and exponential damping times. 
As described above, we obtain $\tau_\mathrm{d}$ and $\tau_\mathrm{g}$ through fitting Eq.~\ref{fit} for simulations with $A \leq 0.01$. For the higher amplitude simulations, we fit the Gaussian of Eq.~\ref{fit} to obtain $\tau_\mathrm{g}$ ($\tau_\mathrm{s} \geq t_\mathrm{f}$) and Eq.~\ref{expon} for $\tau_\mathrm{d}$. We can see that we recover the theoretical value for $t_\mathrm{s}$ well (within 10\%) for $A \leq 0.01$, but for higher amplitudes there are deviations, especially for the transitional amplitude of $A = 0.02$. Once again, this shows that we have to be careful when applying seismology on high amplitude, nonlinear oscillations. \\
Very recently, observational evidence has been produced showing that the damping time of transverse oscillations of coronal loops is a function of the initial displacement amplitude \citep{2016A&A...590L...5G} . Unfortunately, a quantitative comparison of our results to observations is not possible given the large parameter space which would need to be covered. On the other hand, we can state that there are encouraging qualitative agreements between our Figure~\ref{damping} and Figure 2 of \citet{2016A&A...590L...5G}: Firstly, the damping time is decreasing as the initial amplitude of the oscillations increases, as is the case for our simulations with thin inhomogeneous layers. Secondly, this dependence weakens as the amplitude increases. The observational data also indicates that there is a considerable population between the values 1 and 2 of the damping time to the oscillation period ratio, which seem to be weakly dependent on the amplitude, similarly to  our case of the loops with thick inhomogeneous layers. It is interesting to note that the transitional regime (defined earlier) appears to occur for much higher amplitudes for the observed oscillations than in our simulations. This could possibly imply a lower growth rate of the KHI for the observed loops.

\section{Conclusions}

In this study, we investigated the standing kink oscillations of a straight flux tube, aiming to model transverse oscillations of coronal loops. We ran 3D ideal MHD simulations, exploring the parameter space of initial amplitudes and inhomogeneous layer thicknesses. The chosen amplitudes cover the weakly linear and fully nonlinear regimes. The resulting oscillation periods are well described by the linear theory. Furthermore,  for low amplitudes, we find that the resulting damping of the oscillation is close to the damping times computed by linear theory. However, for higher amplitudes, nonlinear effects have a definitive impact on the resulting oscillation characteristics, especially the development of the KHI around the loop edges where the velocity shear is highest, coinciding with the layer where resonant absorption is taking place. The development of KHI (and ultimately the threshold amplitude at which its effects become important) is dictated by its growth rate, which in turn depends on
the ratio of loop radius to length, oscillation amplitude, Alfvén speeds, inhomogeneous layer thickness and numerical dissipation. We show that even for initially thin inhomogeneous layers, the oscillations undergo rapid damping due to the presence of KHI. For high amplitudes (with KHI developing in under one oscillation period), the damping time is almost independent of the initial thickness of the inhomogeneous layer. On the other hand, for thick inhomogeneous layers, the damping time does not seem to depend on the initial amplitude of the perturbation.
To put things in perspective, the highest amplitude perturbation used in our simulations initially displaces the loop by less than its radius, and is thus at the lower boundary of observed flare-related coronal loop oscillation displacements \citep{2002SoPh..206...99A}. Studying the energy distribution in the anti-node cross-section of the loop, we arrive at the conclusion that the kinetic energy in the inhomogeneous layer is converted to plasma internal energy more quickly in the presence of KHI. The increase in average internal energy is less than one percent, thus the energy budget of the wave is not enough to cause any significant temperature change in the 0.9 MK plasma. However, this heating might be significant for prominences \citep{2015ApJ...809...72A}. Even if the dissipated wave energy is not enough to cause significant heating, we show that if the loop is surrounded by hotter plasma, mixing induced by the KHI can increase the average temperature of the loop.  In the presence of KHI, the peak value of average kinetic energy deposited in the inhomogeneous layer is lower than in simulations without KHI. This is a consequence of the accelerated conversion of kinetic to internal energy in the presence of KHI, which cascades energy to smaller scales where it can be dissipated (by numerical dissipation) more efficiently. The disruption of the resonant layer may also contribute to the reduction in the peak value of average kinetic energy, though it is unclear how effective the resulting `patchy' resonant absorption is. \\
According to the present study, it becomes uncertain whether seismology schemes based on the linear theory for the damping rates of coronal loops are valid for high-amplitude, non-linear transverse oscillations.
Using the observed switch between Gaussian and exponential damping profiles of transverse coronal loop oscillations for coronal seismology has recently been suggested. However, because of the nature of nonlinear kink oscillations it is questionable how accurately one can infer parameters such as inhomogeneous layer thickness and density ratio from the observed damping profiles, given that the non-linear damping times are nearly insensitive to them. {However, it is important that future studies also include Gaussian damping profiles in their analyses, as this profile seems to better describe the damping of nonlinear transverse oscillations of coronal loops.} The new observations of amplitude-dependent damping times qualitatively support our conclusions.

\begin{acknowledgements} The authors would like to thank the referee for valuable comments which helped to improve the manuscript. N.M. acknowledges the Fund for Scientific Research-Flanders (FWO-VLaanderen). T.V.D. was supported by an Odysseus grant, the Belspo IAP P7/08 CHARM network and the GOA-2015-014 (KU Leuven). Inspiration for this research was found during ISSI and ISSI-BJ workshops. Visualization was done with the help of VisIt software (\cite{HPV:VisIt}). \end{acknowledgements}

\bibliographystyle{aa} 
\bibliography{../Biblio}{} 

\end{document}